\documentclass{article}

\usepackage{arxiv}

\usepackage[utf8]{inputenc} 
\usepackage[T1]{fontenc}    
\usepackage{hyperref}       
\usepackage{url}            
\usepackage{booktabs}       
\usepackage{amsfonts}       
\usepackage{nicefrac}       
\usepackage{microtype}      

\usepackage{authblk}

\usepackage{xcolor} \usepackage{ulem} \usepackage{changebar}

\usepackage{natbib}
\usepackage{graphicx} 
\usepackage{caption} 
\usepackage{subcaption} 

\usepackage{chngpage}
\usepackage{booktabs}
\usepackage{pdflscape}
\usepackage{afterpage}
\usepackage{capt-of}
\usepackage{adjustbox}
\usepackage{array}
\usepackage{amsmath}

\usepackage{gensymb}
\usepackage{savesym}
\graphicspath{{./}{figs/}}

\usepackage{textcomp}

\title{Where Did They Come From, Where Did They Go. Grazing Fireballs}

\author[1]{P.M. Shober \thanks{patrick.shober@postgrad.curtin.edu.au}}
\author[1]{T. Jansen-Sturgeon}
\author[1]{E.K. Sansom}
\author[1]{H.A.R. Devillepoix}
\author[1]{M.C. Towner}
\author[1]{P.A. Bland}
\author[1]{M. Cup\'ak}
\author[1]{R.M. Howie}
\author[1]{B.A.D. Hartig}
\affil[1]{Space Science and Technology Centre (SSTC), School of Earth and Planetary Sciences, Curtin University, GPO Box U1987, Perth, Western Australia 6845, Australia}

\begin{document}
\maketitle

\bibstyle{abbrvnat}

\begin{abstract}
For centuries extremely-long grazing fireball displays have fascinated observers and inspired people to ponder about their origins. The Desert Fireball Network (DFN) is the largest single fireball network in the world, covering about one third of Australian skies. This expansive size has enabled us to capture a majority of the atmospheric trajectory of a spectacular grazing event that lasted over 90\,seconds, penetrated as deep as $\sim58.5$\,km, and traveled over 1,300\,km through the atmosphere before exiting back into interplanetary space. Based on our triangulation and dynamic analyses of the event, we have estimated the initial mass to be at least 60\,kg, which would correspond to a 30\,cm object given a chondritic density ($3500\,kg\,m^{-3}$). However, this initial mass estimate is likely a lower bound, considering the minimal deceleration observed in the luminous phase. The most intriguing quality of this close encounter is that the meteoroid originated from an Apollo-type orbit and was inserted into a Jupiter-family comet (JFC) orbit due to the net energy gained during the close encounter with the Earth. Based on numerical simulations, the meteoroid will likely spend $\sim200$\,kyrs on a JFC orbit and have numerous encounters with Jupiter, the first of which will occur in January-March 2025. Eventually the meteoroid will likely be ejected from the Solar System or be flung into a trans-Neptunian orbit.
\end{abstract}


\section{Introduction}

\subsection{Reports of Grazing Fireballs}
People have reported witnessing brilliantly long-lasting and bright meteor processions for at least hundreds of years. The 1783 `Great Meteor' was estimated to have traveled $>1600\,km$ through the atmosphere over western Europe \citep{cavallo1784ix}. The `Great Comet of 1860', which was most likely an Earth-grazing fireball over the eastern United States, was accounted for in a painting by American landscape artist Frederic Church entitled `The Meteor of 1860' and by American poet Walt Whitman in his poem `Year of Meteors' \citep{olson2010literary}. Additionally, the `1913 Great Meteor Procession' reported sightings across Canada, the north-eastern United States, Bermuda, and many ships in the Atlantic as far south as Brazil. The event was initially hypothesized to have been formed by a natural Earth satellite that had a grazing encounter with the atmosphere \citep{1913JRASC...7..145C,1916JRASC..10..294D}. 

A grazing event is considered to be when a meteoroid impacts the atmosphere at an extremely low-angle relative to the horizon, and there are generally three possible outcomes. It can either escape back to interplanetary space after passing through the atmosphere, fully ablate, or slow down enough to fall to the Earth. The first scientifically observed and triangulated grazing event was not until 1972 over Canada and the northwestern United States \citep{ceplecha1979earth,ceplecha1994earth}. The 1972 fireball lasted $\sim100\,sec$, covering over 1500\,km, and reached a minimum height of 58\,km. \citet{ceplecha1979earth} estimated the mass to be between $10^{5}-10^{6}$\,kg with the most likely diameter of about 5\,m. The original analysis done by \citet{rawcliffe1974meteor} and \citet{jacchia1974meteorite} is known to contain mistakes, and the values given should not be relied upon \citep{ceplecha1979earth}. 

Since the 1972 fireball, there have been several atmospheric grazing events reported within scientific literature: 
\begin{itemize}
    \item In 1990, \citet{borovicka1992earth} published analysis of the first Earth-grazing fireball observed by a photographic fireball network in which the meteoroid was estimated to be 44\,kg with the closest approach of 98\,km detected by two Czech stations of the European Fireball Network.
    
    \item In October of 1992, a bright fireball endured for over 700\,km over the eastern United States before dropping a meteorite in Peekskill, New York \citep{brown1994orbit,beech1995fall,ceplecka1996video}.
    
    \item In 1996 a fireball was observed to hit the western United States and only briefly escape for one orbit before allegedly impacting the Earth \citep{1997SPIE.3116..156R}.
    
    \item On March 29, 2006, a $\sim40$\,sec grazing fireball was observed over Japan \citep{abe2006earth}. The meteoroid traveled over 700\,km through the atmosphere and reached a minimum height of 71.4\,km. It appeared to come from a JFC-like orbit and the spectra collected was consistent with a chondritic composition. 
    
    \item On August 7, 2007, a grazing fireball was observed by the European Fireball Network originating from a Aten-type orbit \citep{spurny2008precise}.
    
    \item In June 2012, the first grazing meteoroid associated with a meteor shower in the scientific literature was recorded by 13 stations with a 98\,km minimum altitude over Spain and Portugal and belonged to the daytime $\zeta$-Perseid shower \citep{madiedo2016earth}.
    
    \item In 2003, another grazing meteor, mass loss $\approx 5\times10^{-3}$\,g, was detected over Ukraine before exiting back into interplanetary space \citep{kozaketal}.
    
    \item In December 2014, a 1200\,km long grazing event occurred over Algeria, Spain, and Portugal and lasted approximately 60\,seconds, reaching a minimum height of 75\,km \citep{moreno2016preliminary}.
    
    \item On March 31, 2014, a $\sim34$\,sec fireball over Germany and Austria originating from an Apollo-type orbit was observed. The meteoroid was estimated to have an initial mass of about 200\,kg, but no material is believed to have exited back into interplanetary space \citep{oberst2014extraordinary}. Many meteorites may have survived to the ground, however, the uncertainty on the fall ellipse is very large due to the extremely shallow entry angle.
\end{itemize}

For some of these grazing meteoroids mentioned above, the object was able to survive its passage through the atmosphere. The rock then re-entered interplanetary space on an altered orbit, sending material from one part of the inner Solar System to another. This is could be significant since various parts of the inner Solar System are thought to be dynamically and physically distinct from one another.

\subsection{Small Inner Solar System Bodies}
The classical view of the Solar System says that the Sun formed with a debris disk around it that was originally compositionally heterogeneous within bands of constant radial distance from the Sun. The `snow line' denoted the boundary between the planetesimals in which water ice and other volatiles would be retained and the bodies which were unable to hold ice, thus remaining dry. This classically separated the small bodies within the Solar System into two main groups: comets and asteroids respectively. 

Although, we have seen that this classical ideology does not usually fit our observations of the small bodies within the Solar System. The Solar System is complicated and dynamic. In the last 4.5\,billion years, small bodies have been jumbled around and altered. The layout and distribution of the Solar System is much more complicated than the idealized stratified one we tend to imagine. 

Within the inner Solar System there are short-period comets, main-belt objects (MBOs), and near-Earth objects (NEOs). Traditionally, the MBOs were considered asteroidal and inner Solar System in origin, and NEOs primarily evolved from the MBO space after entering an orbital resonance \citep{bottke2002debiased,granvik2018debiased}. However, with the identifications of Main-Belt Comets (MBCs) \citep{hsieh2006population} and dry asteroidal material in the Kuiper Belt \citep{meech2016inner}, we have realized that the material in the Solar System is more mixed than previously believed \citep{fernandez2015jupiter}. Additionally, the starkly drawn lines between asteroidal and cometary material have since faded with the identification of active asteroids, extinct comets, and mixing between populations \citep{fernandez2001low,fernandez2002there,fernandez2005albedos,kim2014physical,jewitt2012active}. In reality, the physical properties of small bodies in the Solar System most likely exist in a spectrum from primitive volatile-rich (``comet-like") to dry volatile-poor (``asteroid-like"). We are still trying to determine the most probable mechanism by which this mixing could have occurred, but several models such as the `Nice Model' and the `Grand Tack' have begun to elucidate some of these mysteries \citep{walsh2011low,tsiganis2005origin}.

Jupiter family comets (JFCs) are a class of short-period comets, believed to have evolved from scattered disk and Kuiper belt orbits \citep{fernandez1980existence,levisonduncan1997,duncan1997disk,binzel2004observed}. JFCs are primitive and contain a large amount of hydrated minerals and volatile ices \citep{kelley2009composition,jenniskens2012radar}. They are also characterized by their orbits being strongly linked to the orbit of Jupiter, typically defined by their Tisserand's parameter to be $2 < T_{J} < 3$ \citep{carusi1987dynamical,levison1994long}. JFCs usually have multiple low-velocity encounters with the gas-giant over their lifetime \citep{levisonduncan1997,duncan1997disk,duncan2004dynamical}. These encounters with Jupiter make the orbits of JFCs more unpredictable compared to other small bodies, where the median dynamic lifetime of a JFC $\sim 325$\,kyr \citep{duncan2004dynamical}. However, as described by \citet{fernandez2015jupiter}, JFCs that display cometary features frequently encounter Jupiter at distances of $\leq0.1$\,AU making them highly unstable compared to a small subset of near-Earth ``asteroidal" JFCs which typically exist on more stable orbits comparatively. A tiny fraction of JFCs are also thought to decouple from Jupiter and become Encke-like comets through either non-gravitational perturbations or close planetary encounters \citep{steel1996origin,levison2006origin}. 


\subsection{The Desert Fireball Network}
Since 2003, the Desert Fireball Network (DFN) has been operating observatories across south-western Australia to capture images of fireball events \citep{bland2004desert}. The network has since grown from 4 observatories by 2007 to over 50 observatories in Western Australia and South Australia by 2015 \citep{bland2012australian,howie2017build}. No other fireball camera network in the world is this expansive. Furthermore, we have expanded this effort worldwide with the start of Global Fireball Observatory (GFO) collaboration (Devillepoix et al., in prep.) with coverage area expected to increase to 2\% of the Earth's entire surface. This coverage area makes the GFO particularly well suited to characterize grazing meteoroids and other more rare fireball events \citep{shober2019identification}.


\section{DFN Observations}

\begin{figure}
	\centering
	\includegraphics[width=\textwidth,height=\textheight,keepaspectratio]{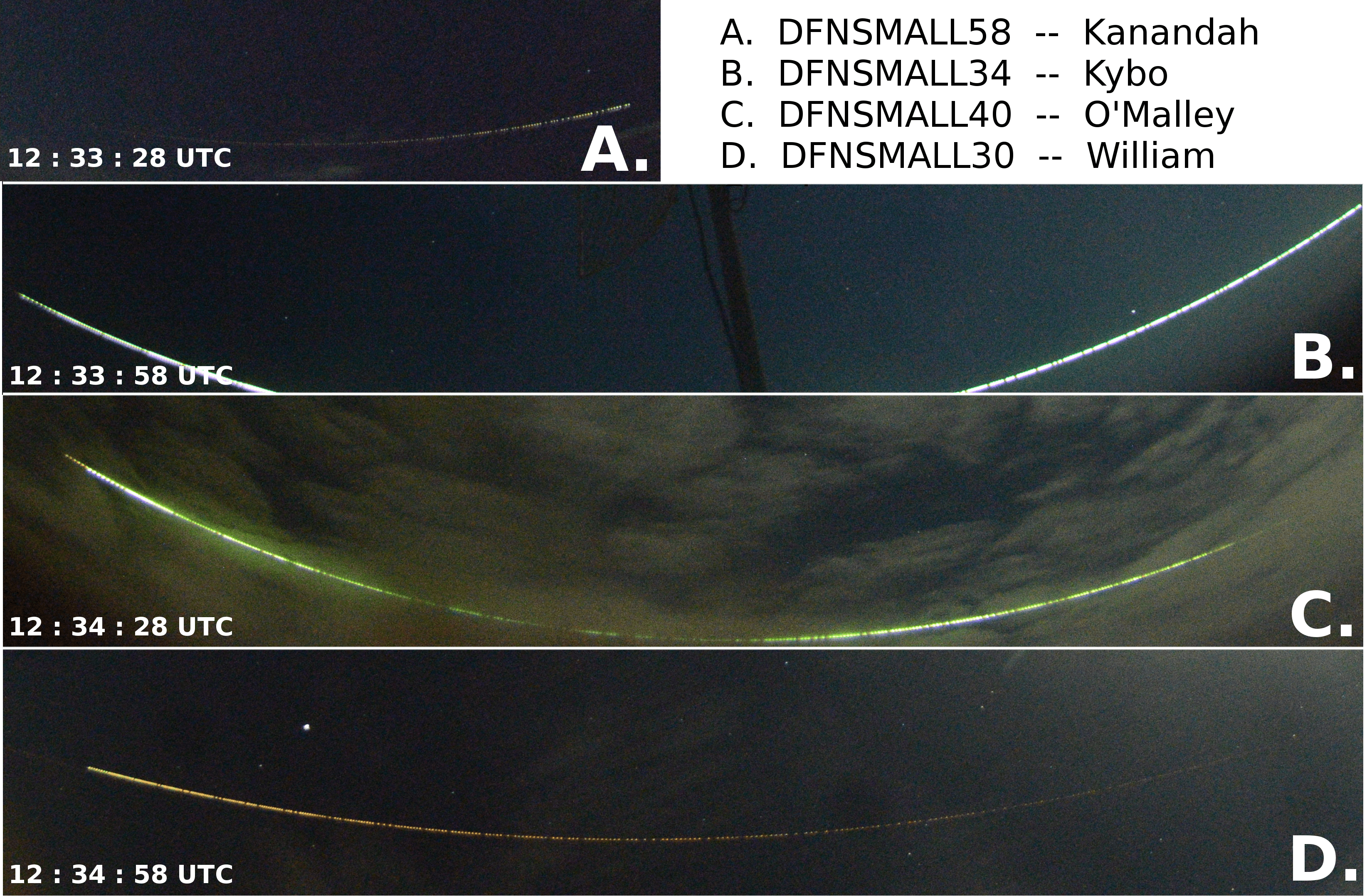}
	\caption{Long exposure images of event DN170707\_01. The event lasted over 90\,seconds and spanned four 30\,second exposures (A, B, C, D). The fireball was first observed at 85\,km altitude, reached as low as 58\,km, and then was visible until 86\,km before escaping the Earth's atmosphere. The initial velocity was $16.1\,\mbox{km s}^{-1}$, and the exit velocity after passing through the atmosphere was about $14.6\,\mbox{km s}^{-1}$. The images are all oriented so that the fireball travels from left to right (west to east).}
	\label{fig:cam_images17}
\end{figure}

On July 7th, 2017, a 90\,second extremely shallow fireball was observed to graze the atmosphere above Western Australia and South Australia, entering the atmosphere at a slope of $\sim4.6\degree$ (Fig.~\ref{fig:cam_images17}). Ten DFN observatories made observations of the fireball as it traveled over 1300\,km through the atmosphere. The luminous phase started at about 85\,km and penetrated as deep as 58\,km before ceasing to be visibly ablating at 86\,km. This event is only equaled by the `Great Daylight Fireball of 1972', which reached a similar depth and lasted $\sim9$\,seconds longer than our witnessed event \citep{ceplecha1979earth}. However, unlike the 1972 event, the DFN was able to photographically image a majority of the the atmospheric trajectory of the fireball (including the beginning and the end), with observations from many of our fireball observatories spread across Western Australia and South Australia. Thus, providing us with a substantial amount of data to accurately fit a trajectory to our observations (2541 astrometric datapoints). A summary of the observations made of event DN170707\_01 and the fitted trajectory are provided in Table~\ref{tab:obs}. The number of observations refers to the number of 30-second exposures. Whereas, `without timing' denotes when observations of the fireball were collected, however, either the angular velocity of the meteoroid was too slow or the fireball was too bright to distinguish the encoded de Bruijn sequence \citep{howie2017submillisecond}. Unfortunately, due to the DFN's viewing geometry at the beginning of the observed luminous trajectory, the initial observation convergence angle is only a few degrees. Therefore, the uncertainty associated with the initial velocity is higher than usual, however, still sufficient to determine what part of the Solar System the meteoroid originated.

\begin{figure}
	\centering
	\includegraphics[width=\textwidth,height=\textheight,keepaspectratio]{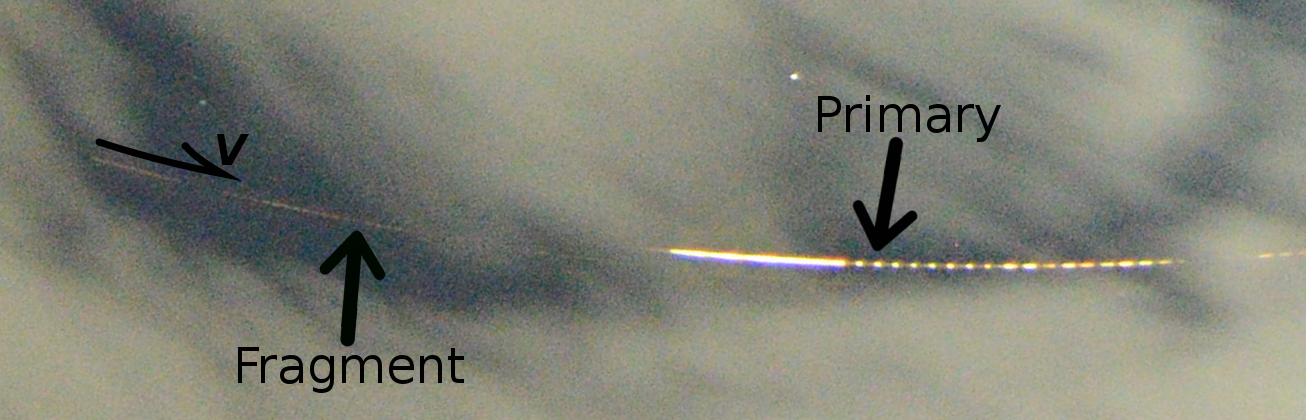}
	\caption{Fragmentation event captured for event DN170707\_01 near the closest approach of its trajectory. The image shows two distinct paths offset from each other. The brighter path on the right side of the image belonging to the primary piece, whereas on the left the trail of a smaller fainter fragment can be seen. The decrease in velocity due to the observed fragmentation was not significant relative to the velocity scatter, and thus was not included during the trajectory fit. Additionally, only one camera observed the fragment due to cloud coverage and geometry, and therefore a trajectory for the fragment was unable to be determined. No other fragmentation events were detected along the path.}
	\label{fig:fragment}
\end{figure}

\begin{figure}
	\centering
	\includegraphics[width=\textwidth,height=\textheight,keepaspectratio]{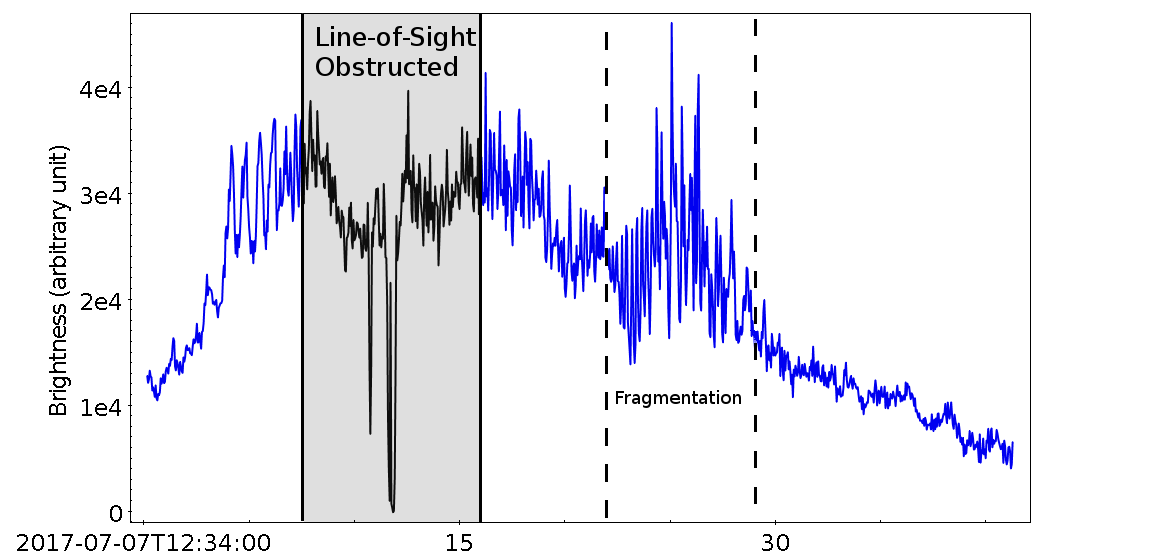}
	\caption{Light curve based on video from observatory DFNSMALL34-Kybo in the Nullarbor Plain in Western Australia during the fragmentation of event DN170707\_01. The fragmentation occurs about 25\,sec into this exposure (enclosed by dashed vertical lines), forming one detectable fragment. The y-axis is brightness in arbitrary units due to the photometry data lacking calibration. The ram pressure on the meteoroid just prior to the fragmentation was $\sim0.08$\,MPa. The line-of-sight was briefly obstructed by a telephone pole, reducing the brightness.}
	\label{fig:fragcurve}
\end{figure}

At the meteoroid's closest approach, a fragmentation event occurred in which a smaller piece of the primary object broke off (Fig.~\ref{fig:fragment}). DFN observatories captured the fragmentation event on video, and an uncalibrated light curve was able to be extracted (Fig.~\ref{fig:fragcurve}). There are no other instances of fragmentation observed during the trajectory. This fragmentation event was taken into account when triangulating the path of the primary and determining the mass of the meteoroid. 

\begin{table}
\centering
\caption{Observations and triangulated trajectory for event DN170707\_01, recorded over Western Australia and South Australia on July 7th, 2017. Mass range determined by varying density between $2800-7300\ kg\ m^{-3}$ and includes formal uncertainties. Timing uncertainty is nominally $10^{-4}-10^{-5}$\,s, considerably less than other sources of uncertainty for the trajectory \citep{howie2017submillisecond}.}
\begin{tabular}{| l  c  c |}
\hline\hline
                                        &     Entry Conditions       &    Exit Conditions        \\ 
\hline\hline 
Time (UTC) after 2017-07-07             &  12:33:45.900              & 12:35:16.050              \\
Height (km)                             &  $85.66\pm0.03$            & $86.04\pm0.02$            \\ 
Mass range (depending on density; kg)   &  14-92                     & 9-62                      \\ 
Latitude (deg)                          &  $-28.6933\pm0.0003$       & $-28.4144\pm0.0002$       \\ 
Longitude (deg)                         &  $122.7161\pm0.0010$       & $136.3318\pm0.0002$       \\ 
Velocity ($\mbox{km\,s}^{-1}$)          &  $15.71\pm0.13$            & $14.24\pm0.10$            \\
Slope (deg)                             &  $4.6$                     & $7.8$                     \\ 
\hline\hline
Duration (sec)                          & \multicolumn{2}{c|}{90.15}                             \\ 
Minimum Height (km)                     & \multicolumn{2}{c|}{58.5}                              \\
Best Convergence Angle (deg)            & \multicolumn{2}{c|}{45.9}                              \\ 
Number of Observations (with timing)    & \multicolumn{2}{c|}{13}                                \\ 
Number of Observations (without timing) & \multicolumn{2}{c|}{7}                                 \\ 
Number of Datapoints                    & \multicolumn{2}{c|}{2541}                              \\ 
\hline
\end{tabular}
\label{tab:obs}
\end{table}

\section{Methods}

\subsection{Triangulation}
In the past, fireball and meteor observation networks estimated the trajectories they witnessed using a simplified straight-line-fit approach \citep{ceplecha1987geometric,borovicka1990comparison}. These simplified straight-line fit techniques are sufficient enough to obtain meaningful results when the trajectory is shorter than 100\,km. However, recent studies have shown that more satisfactory results can be obtained with the use of more rigorous methodologies \citep{sansom2015novel,sansom20193d,jansen2019}. This is particularly true for a grazing fireball where the meteoroid is traveling hundreds to thousands of kilometers through the atmosphere. In previous grazing fireball studies, this non-linearity was accounted for in several different ways. \citet{ceplecha1979earth} was the first to recognize that a grazing trajectory should fit a hyperbola when neglecting the atmosphere, but is otherwise slightly more curved due to the atmospheric drag experienced. Thus, \citet{ceplecha1979earth} fit osculating circles to the trajectory of the 1972 grazing daylight fireball to account for this added curvature with reasonable accuracy. \citet{borovicka1992earth} utilized the fact that one of the observation stations was nearly directly below the fireball (passed nearly through zenith) and saw the entire trajectory. They took their observations and performed a least-squares fit to an osculating circle at the point of pericenter, neglecting drag in this case based on fireball type. Similar methodologies using osculating circular trajectory fits have been utilized by other studies as well \citep{abe2006earth}. \citet{kozaketal} triangulated a small, fast grazing, high-altitude meteor detected by video observatories in Ukraine by assuming minimal drag and fitting the observations to a hyperbolic orbit in the geocentric frame. Meanwhile, \citet{madiedo2016earth} determined the atmospheric trajectory of a meteor belonging to the Daytime $\zeta$-Perseid shower by using a segmented method-of-planes approach adapted from \citet{ceplecha1987geometric}.

For standard DFN events, we implement a modified straight-line least-squares (SLLS) method with an Extended Kalman Smoother (EKS) for velocity determination \citep{sansom2015novel}. We then numerically determine the meteoroid's orbit by including all relevant perturbations. Numerical methods are a slightly more accurate way to handle the orbit determination, especially for meteoroids that were slow or closely approached the Moon \citep{clark_weigert2011,dmitriev2015orbit,jansen2019}. For longer and/or shallower fireball events, where the meteoroid trajectory can have noticeable curvature, the SLLS method cannot account for the non-linear motion. Within this study, we implemented a Dynamic Trajectory Fit (DTF) triangulation method that fits the observation rays directly to the equations of motion for fireballs \citep{jansensturgeon2019dynamic}. This non-straight-line approach to the event triangulation represents the physical system more veraciously. Consequently, the DTF method produces a much better fit to the observations compared to the SLLS for both positions and velocities (Fig.~\ref{fig:height_time} \& Fig.~\ref{fig:vel_time}). We then use this trajectory (Fig.~\ref{fig:LS_triang17}), to numerically estimate the pre- and post-grazing orbits. Although, currently the DTF method does not provide adequate formal velocity errors, thus a EKS was utilized to determine the velocity uncertainties for this study. 

\subsection{Mass Determination}
During the DTF procedure, the meteoroid's ballistic parameter and ablation coefficient are determined alongside its dynamic parameters, based directly on the line-of-sight observations. By assuming the meteoroid's shape and density, a mass estimation can be deduced from the meteoroid's fitted ballistic parameter.

\subsection{Orbital Integration}
After triangulating the grazing event, we initialized several orbital integrations using the publicly available REBOUND code\footnote{http://github.com/hannorein/REBOUND} \citep{2012A&A...537A.128R}. We utilized the 15th order non-symplectic IAS15 integrator for our simulations of the event \citep{2015MNRAS.446.1424R}. This integrator is based upon the RADAU-15 integrator developed by \citet{1985dcto.proc..185E}. It improves upon its predecessor by minimizing the systematic error generated by the algorithm to well-below machine precision, implementing an adaptive time-step, and adding the ability to include in non-conservative forces easily while ensuring that the round-off errors are symmetric and at machine-precision. 

\paragraph{Initialization}
From the trajectory determined by the DTF method, the pre- and post-atmospheric state vectors for the meteoroid can be used to initialize orbital simulations. These simulations contain N number of particles within the meteoroid state's uncertainties produced by the triangulation. Currently, the DTF methodology does not provide formal uncertainties as model errors are not accounted for \citep{jansensturgeon2019dynamic}. Subsequently, for this event, we determined the velocity uncertainties using the EKS method in conjunction with the DTF trajectory fit. Additionally, we assume a Gaussian distribution for the errors, although this may not be strictly true. However, the results from the integration should not deviate significantly due to this assumption. The particles' positions are generated from the initial and final latitude, longitude, and height determined from the DTF triangulation. The speed of the particles and their right ascension and declination are given in the Earth-centered Earth-fixed (ECEF) frame and then converted to the Earth-centered inertial (ECI) frame in order to generate the particles in the simulations.  

\paragraph{Integration}
Initial simulations were run within $\pm100$ years of the grazing event in order to accurately characterize the short-term evolution of the meteoroid. The number of outputs recorded was increased so that any close encounters with Jupiter or the Earth would be well resolved. Afterward, a series of long-term integrations were done in a similar manner. The primary goal of these more extended integrations was to determine what were the lasting effects of the meteoroid's grazing encounter with the Earth. Does it stay on a JFC orbit as long as any typical JFC, and where does it evolve to after? Each integration recorded the positions, velocities, and osculating orbital elements for the meteoroid particles for a total period of 500,000\,years forward relative to the event epoch. Close encounters with other planets were also considered and inspected, particularly with Jupiter.  


\section{Results and Discussion}

\begin{figure}
    \centering
    \includegraphics[width=\textwidth]{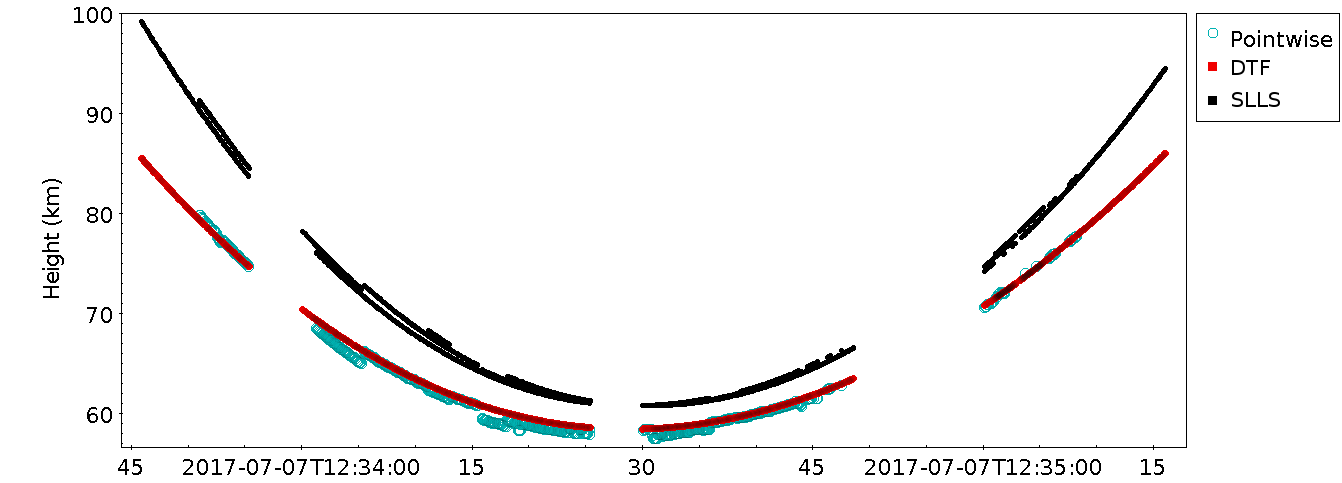}
    \caption{Height variation as a function of time determined by the straight-line least-squares (SLLS) and the Dynamic Trajectory Fit (DTF) methods. The pointwise heights represent the points that minimizing all the angular distances between the simultaneous lines-of-sight (given $> 2$), the respective observatory, and the point itself. The DTF fits much better to the pointwise than the SLLS due to its incorporation of gravity, drag, and ablation. This non-straight line fit produces a much more useful model to understand these grazing fireball events. The shape of the trajectory is somewhat misleading, as the trajectory would be concave with respect to a global, inertial reference frame instead of convex, as shown here. The three distinct gaps in the trajectory are due to latency between observation periods \citep{howie2017build}. This lapse in observations occurs once every thirty seconds and is only typically noticeable for the longest fireball trajectories observed by the DFN. Towards the end of the trajectory, the largest lapse in observations was also due to the cloud coverage at the time}
    \label{fig:height_time}
\end{figure}

\begin{figure}
    \centering
    \includegraphics[width=\textwidth]{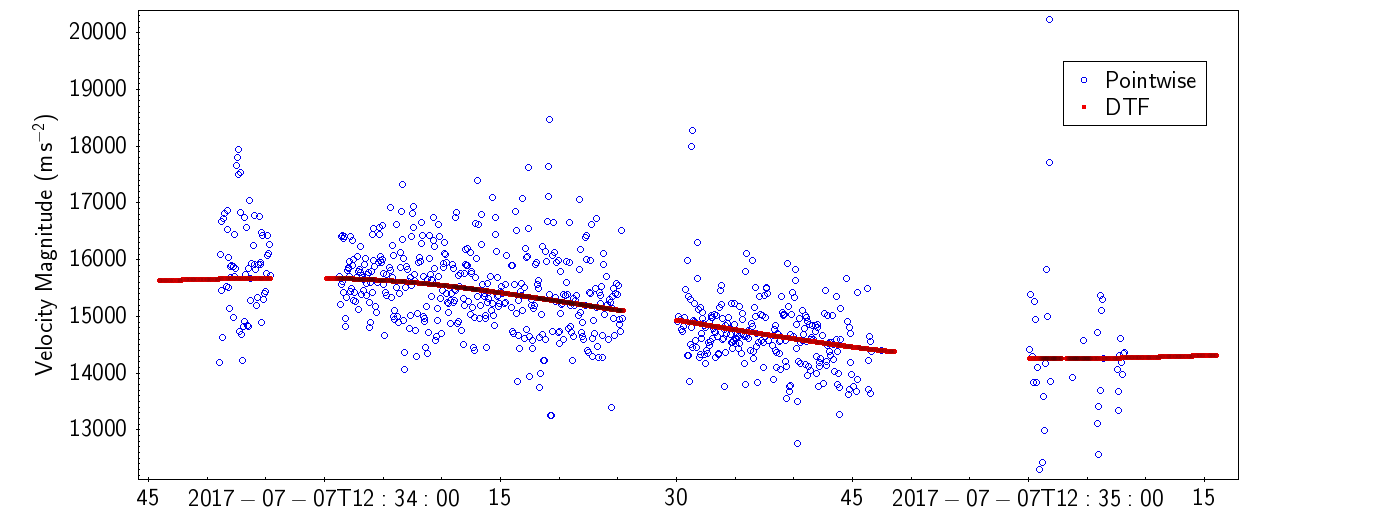}
    \caption{The velocity of the DN170707\_01 meteoroid event as determined using both the Dynamic Trajectory Fit model (red) and a pointwise triangulation fit (blue). The DTF method fits the line-of-sight observations directly to the dynamic equations of motion that describe the motion of fireballs. Pointwise scattered instantaneous speeds correspond to the center-difference between adjacent data points seen by $> 2$ observatories. These points in 3D space are calculated by minimizing all the angular distances between the simultaneous lines-of-sight, the respective observatory, and the point itself.}
    \label{fig:vel_time}
\end{figure}

\begin{figure}
	\centering
	\includegraphics[width=\textwidth,height=\textheight,keepaspectratio]{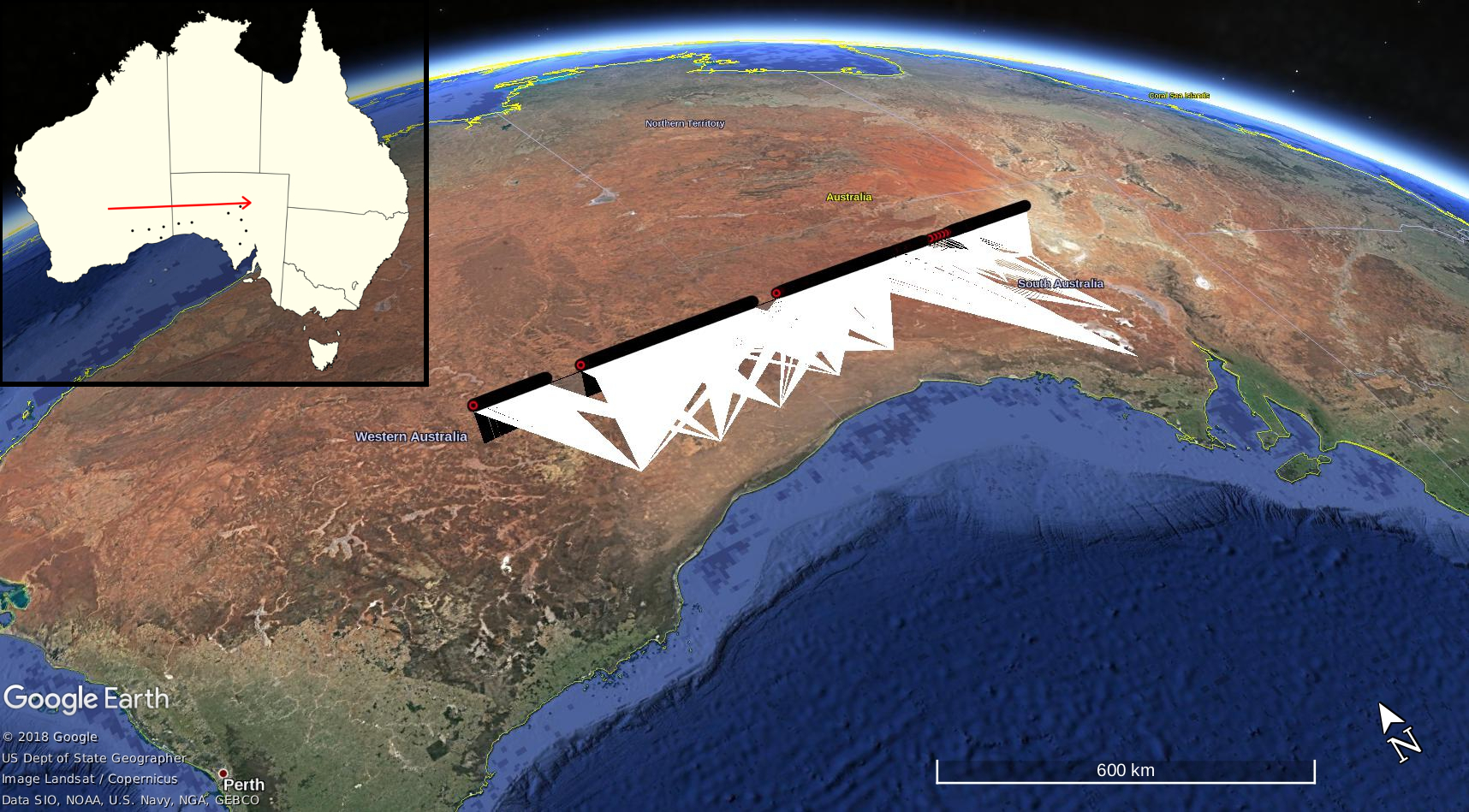}
	\caption{Triangulated luminous atmospheric trajectory for event DN170707\_01, as seen over Western Australia and South Australia. The triangulation method used involves fitting the line-of-sight observations directly to the meteoroid's dynamic equations of motion, thereby dropping any straight-line assumptions \citep{jansensturgeon2019dynamic}. The event lasted 90\,seconds, initially hitting the atmosphere at $4.6\degree$ and covering over 1300\,km through the atmosphere. The white rays indicate the line-of-sight measurements from each DFN observatory, whereas the black path marks the triangulated trajectory based on the observations of the fireball.}
	\label{fig:LS_triang17}
\end{figure}

\subsection{Atmospheric Trajectory}
As seen in Fig.~\ref{fig:height_time}, the DTF methods fit the pointwise observations much better than the SLLS method for an event that is thousands of kilometers in length. The pointwise heights are given by minimizing the angular distance between the lines-of-sight when at least two observations are made. If a center-difference is taken between all these points, a velocity scatter can be generated (Fig.~\ref{fig:vel_time}). The velocity scatter for event DN170707\_01 is very large in some circumstances considering the low convergence angles especially for the beginning of the trajectory. A majority of the fireball's trajectory was north of the DFN observatories (Fig.~\ref{fig:LS_triang17}). Thus reducing the accuracy of each measurement. However, since we gathered over 2500 datapoints from ten DFN observatories, a reasonably good trajectory was able to be extracted. There are also three distinct gaps in the observations of event DN170707\_01 primarily due to the latency between the 30-sec observation periods. These lapses in observations are typically only noticeable for the longest enduring fireballs observed by the DFN. The longest gap, towards the end of the trajectory, is compounded by the poor visibility for the DFN observatories in that area of the network due to the cloud coverage at the time. 

\begin{figure}
	\centering
	\includegraphics[width=\textwidth,height=\textheight,keepaspectratio]{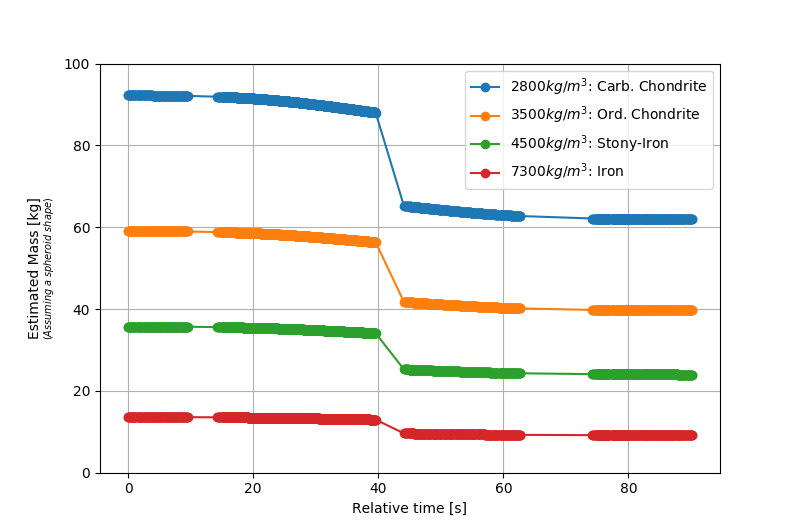}
	\caption{Mass estimation based on DTF triangulation fit to the DFN's observations. The fragmentation event was taken into account, as seen by the sudden mass loss experienced at $\sim40$\,sec into the luminous phase. Each line represents a different density estimate for the object, given the DTF ballistic parameter.} 
	\label{fig:mass_plot}
\end{figure}

During the DTF procedure, the ballistic parameter was determined throughout the trajectory based directly on the line-of-sight measurements, and hinges on the deceleration profile of the observed meteoroid. The meteoroid's mass was estimated by assuming its shape and density, as seen in Fig.~\ref{fig:mass_plot}. For instance, assuming a spheroid of chondritic density ($3500\,kg\,m^{-3}$), the DN170707\_01 meteoroid was estimated to have a 60\,kg initial mass and a 40\,kg outbound mass. A majority of the mass loss is predicted to have occurred during the fragmentation observed near the closest approach of the object. However, as minimal deceleration was observed during the luminous atmospheric encounter (Fig.~\ref{fig:vel_time}), this mass estimate would be more accurately viewed as a lower bound.

The loading ram pressure for the meteoroid at the time of fragmentation was also calculated using the following equation: 
\begin{equation} \label{eq:ram_pressure}
    p = \rho_{h}v_{h}^2
\end{equation}
where $\rho_{h}$ is is the atmospheric density at the height $h$ of the fragmentation and $v_{h}$ is the speed of the meteoroid at that instant. For event DN170707\_01, we determined the fragmentation height based on the time of fragmentation observed in the light-curve from video observations. We estimated the meteoroid to have fragmented at $58.49\pm0.01$\,km, just before the minimum height reached, with a velocity of $15.5\pm0.1$\,km\,$s^{-1}$. We then used the NRLMSISE-00 global atmospheric model to determine the density of the atmosphere at the fragmentation height \citep{picone_nrlmsise_00}. The ram pressure experienced by the meteoroid just before fragmentation was calculated to be $0.084\pm0.01$\,MPa. This very low-value is consistent with the results of \citet{popova2011very}, in which it was found that bulk strengths determined by initial fragmentation are consistently much lower than the strengths of recovered meteorites. Thus, this value likely reflects macro-scale fractures in the object and not the intrinsic material strength. For example, the Dingle Dell ordinary chondrite meteorite recovered by the DFN in 2016 also experienced similar low-pressure fragmentations ($0.03-0.11$\,MPa) early in its brightflight, despite having a recovered bulk density of $3450\,\mbox{kg\,m}^{-3}$ \citep{devillepoix2018dingle}.

\begin{figure}
	\centering
	\includegraphics[width=\textwidth,keepaspectratio]{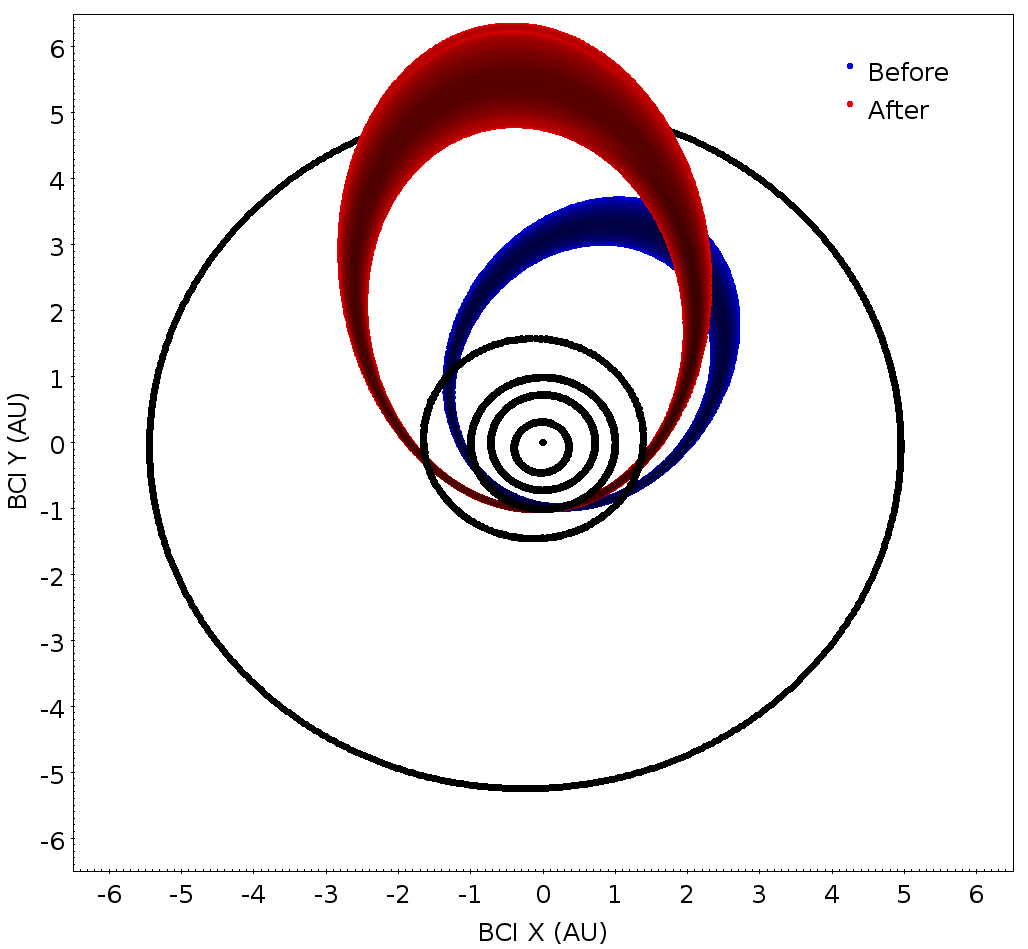}
	\caption{The meteoroid's orbit before and after the grazing encounter with the Earth. The meteoroid originated from an Apollo-type asteroidal orbit and was inserted into a JFC orbit. Once in this JFC orbit, the object's path rapidly becomes less certain due to multiple close-encounters with Jupiter.}
	\label{fig:before_after}
\end{figure}

\subsection{Short-term Simulations}
\begin{figure}
	\centering
    \includegraphics[width=\textwidth,keepaspectratio]{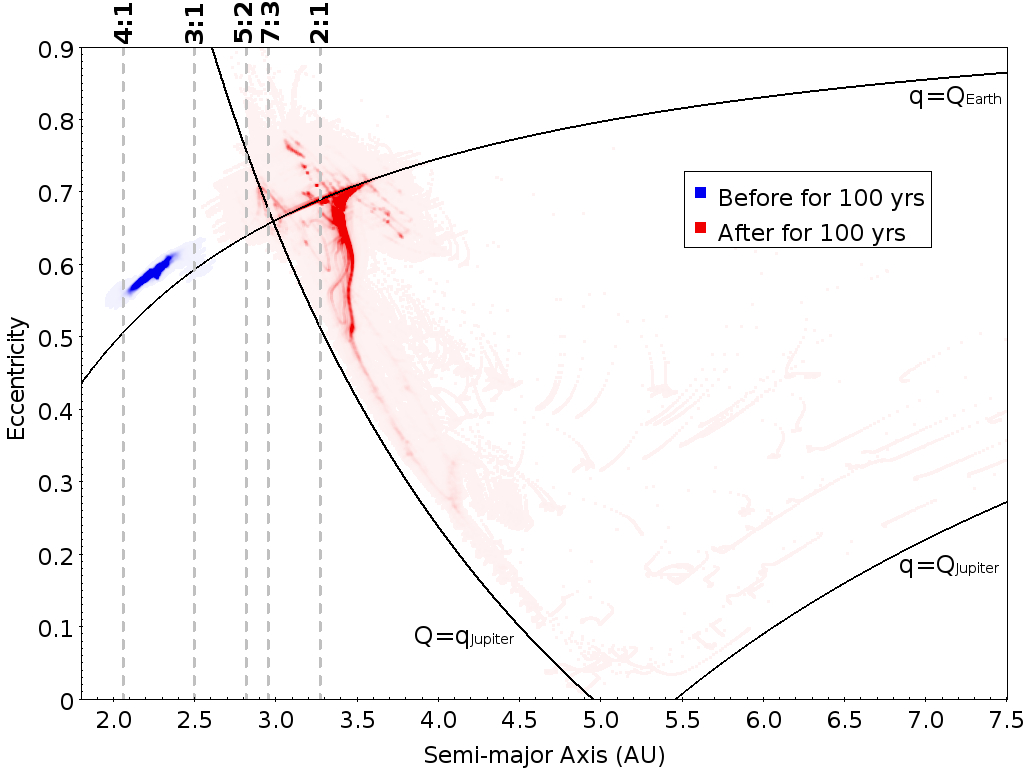}
	\caption{Semi-major axis vs. eccentricity during $\pm100$ years of integrations involving $10,000$ test particles. Particle density over time is indicated by opacity. A majority of the particles remain close together after the grazing encounter, with a small number of particles being scattered by Jupiter very quickly. The significant mean-motion resonances are also plotted as vertical dotted lines. The object came from an eccentric orbit between the 4:1 and 3:1 mean motion resonances. After the grazing encounter with the Earth, the object gained energy and was transferred onto a JFC orbit near the 2:1 resonance with Jupiter. In this orbit, the future of the meteoroid is strongly influenced by the gas giant. Over time, the meteoroid will tend to follow the aphelion and perihelion lines for Jupiter.}
	\label{fig:pm100}
\end{figure}

\begin{figure}
	\centering
    \includegraphics[width=\textwidth,keepaspectratio]{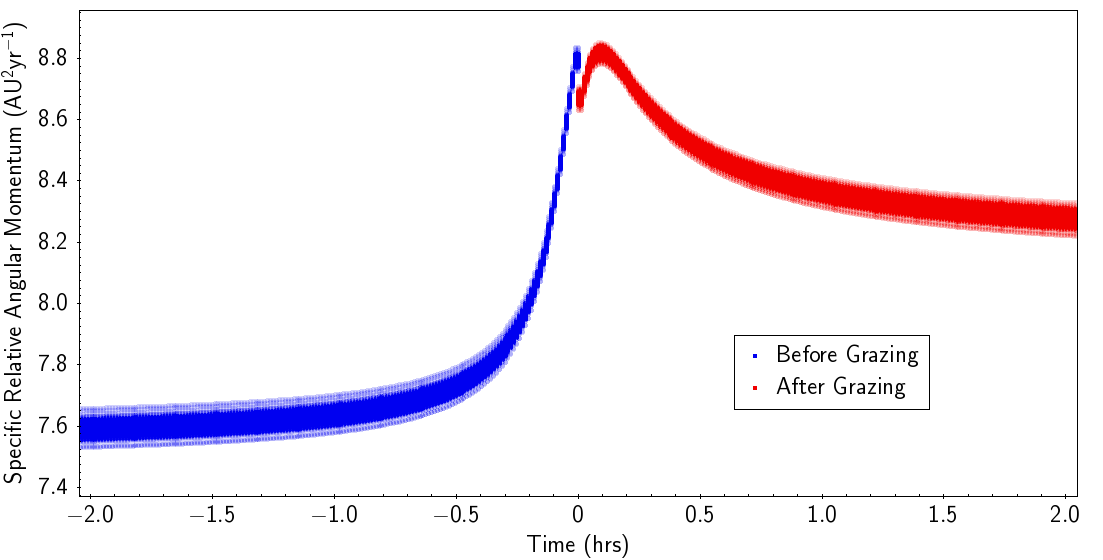}
	\caption{Specific relative angular momentum of the meteoroid $\pm12$ hours relative to the grazing event. The meteoroid gains energy after its encounter with the Earth despite losing some energy during the atmospheric passage. At time~=~0, the discontinuity is due to the exclusion of the time when the meteoroid was passing through the atmosphere. The `instant' drop in energy here corresponds to the energy lost due to atmospheric drag. The object continues to gain angular momentum briefly after leaving the atmosphere before losing some energy as it travels away from the Earth. This net gain in angular momentum effectively increased the semi-major axis and eccentricity of the body.}
	\label{fig:ang_m}
\end{figure}

As shown in Table~\ref{tab:before_after}, the meteoroid that skipped off the atmosphere over Western Australia and South Australia in July 2017 originally came from an orbit in the inner main-belt, between the 4:1 and the 3:1 mean-motion resonances with Jupiter (Fig.~\ref{fig:pm100}). It most likely evolved into an Earth-crossing orbit after passing through either the 3:1 or the $\nu_{6}$ complex, which are the two most significant entry routes into the NEO region \citep{bottke2002debiased,granvik2018debiased}. As a result of the grazing encounter with the Earth, the meteoroid was flung into an orbit with a higher energy (Fig.~\ref{fig:before_after}). The geometry of the encounter enabled the meteoroid to gain angular momentum around the Sun (Fig.~\ref{fig:ang_m}). As a result, the semi-major axis and eccentricity both increased due to the increase in energy, and the object was inserted into a JFC orbit. Hereon, the object's future is strongly governed by its interactions with the gas-giant. Fig.~\ref{fig:pm100} shows the evolution of the orbital elements for the meteoroid $\pm 100$ years relative to the grazing encounter. 

With an post-ecounter aphelion near Jupiter's orbit, the meteoroid is likely to have multiple close-encounters with the planet in the future. Thus, the object is unpredictable on relatively short timescales compared to other small bodies in the Solar System. This is to be expected for an object on a JFC-like orbit that originated from the trans-Neptunian region \citep{fernandez2015jupiter}. As seen in Fig.~\ref{fig:pm100}, the object will tend to decrease in eccentricity and slightly increase in semi-major axis over time. This will occur slowly for a majority of particles over about 10-100\,kyrs, as Jupiter perturbs them. If the orbit of the meteoroid evolves into an orbit with a similar semi-major axis to Jupiter, the close encounters with the gas giant will begin to increase the eccentricity of the meteoroid again and throw the body towards the outer Solar System. The meteoroid is also nearly centered on the 2:1 mean motion resonance (Fig.~\ref{fig:pm100}), however, this resonance is not as destabilizing as the other prominent resonances on such short timescales \citep{morbidelli2002origin}.

\begin{table}[]
    \centering
    \begin{tabular}{| l | c | c |}
    \hline\hline
                    &        Before      &       After       \\
    \hline\hline 
     a (AU)         & $2.23\pm0.06$      & $3.26\pm0.12$     \\
     e              & $0.59\pm0.01$      & $0.69\pm0.01$     \\
     $i$ (deg)      & $2.79\pm0.04$      & $3.30\pm0.04$     \\
     $\Omega$ (deg) & $286.46\pm6.03$    & $285.29\pm0.01$   \\
     $\omega$ (deg) & $316.43\pm3.56$    & $350.91\pm0.29$   \\
     q (AU)         & $0.9104\pm0.0003$  & $1.007\pm0.0004$  \\
     Q (AU)         & $3.458\pm0.114$    & $5.36\pm0.2300$   \\
     T$_{J}$        & $3.41\pm0.05$      & $2.75\pm0.05$     \\
    \hline\hline
    \end{tabular}
    \caption{Heliocentric orbital elements for the meteoroid associated with event DN170707\_01 just before and after its grazing encounter with the Earth. The uncertainties of the orbital elements were determined by a short Monte Carlo simulation consisting of 5,000 particles randomly generated within triangulation errors and numerically integrated forward and backward relative to the grazing event. The immediate effect of the encounter on the orbit is apparent; the semi-major axis, eccentricity, and argument of perihelion of the meteoroid were all significantly increased. The grazing encounter changed the orbit of the meteoroid from an Apollo-type NEO to a JFC orbit. The resulting orbit is comparatively unstable due to its aphelion being very similar to the semi-major axis of Jupiter, increasing the chance of a close encounter with the gas giant.}
    \label{tab:before_after}
\end{table}

\begin{figure}
    \begin{subfigure}[b]{\textwidth}
         \centering
         \includegraphics[width=\textwidth]{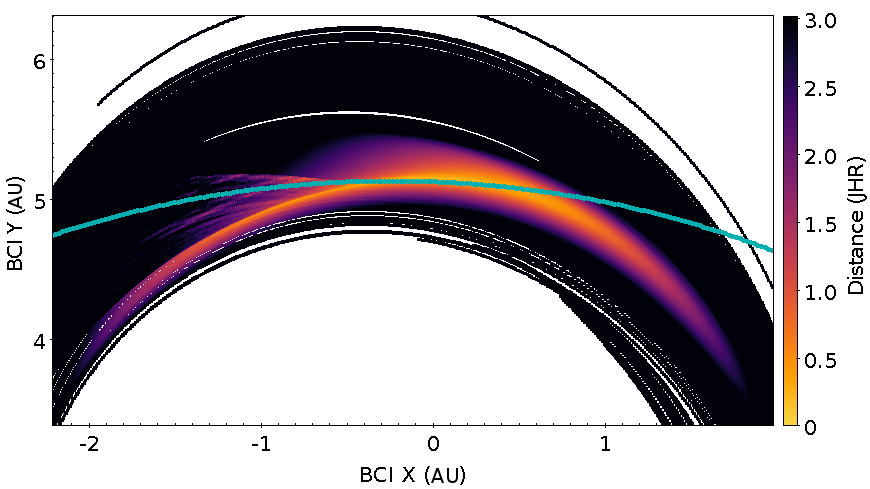}
         \caption{Close encounter of meteoroid with Jupiter (blue path).}
         \label{fig:cesub1}
     \end{subfigure}
     \\
     \begin{subfigure}[b]{\textwidth}
         \centering
         \includegraphics[width=\textwidth]{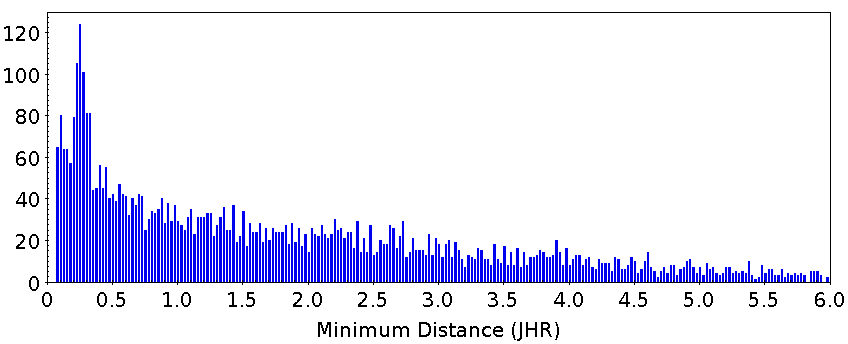}
         \caption{Histogram of the minimum distances between the particles and Jupiter.}
         \label{fig:cesub2}
     \end{subfigure}
    \caption{After grazing Earth's atmosphere, the meteoroid will complete 1.5 orbits around the Sun before likely having its first close encounter with Jupiter. Both plots provide the distance from Jupiter in terms of Jupiter Hill Radii (JHR). Of the 5,000 particles in this integration, nearly $40\%$ come within 1\,JHR and $80\%$ are within 3\,JHR. The mean approach is about 0.7\,JHR. Consequently, the orbit of the meteoroid is highly uncertain after this point, approximately 7.52\,years after its grazing encounter with the Earth (January-March 2025).}
    \label{fig:ce_fig}
\end{figure}

The first of these close-encounters will most likely occur between January and March, 2025 ($\sim7.52$\,years after encountering the Earth) in which the meteoroid will very likely come within 3 Jupiter Hill radii (JHR) of the planet. A series of short-term highly resolved integrations were performed with 5,000 test particles to analyze this first close encounter with Jupiter. As shown in Fig.~\ref{fig:ce_fig}a, the meteoroid is likely to get close to Jupiter (blue path), just $1.5$ orbits after our observations of the fireball. Fig.~\ref{fig:ce_fig}b shows the minimum distances reached by every particle in the integration, many of which ($40\%$) approaching within 1\,JHR with the mean approach of all particles being $0.7$\,JHR. After this close encounter, the test particles disperse relatively quickly, and precisely predicting the future orbit of the meteoroid becomes unrealistic. 

\begin{figure}
    \begin{subfigure}[b]{\textwidth}
         \centering
         \includegraphics[width=0.45\textwidth]{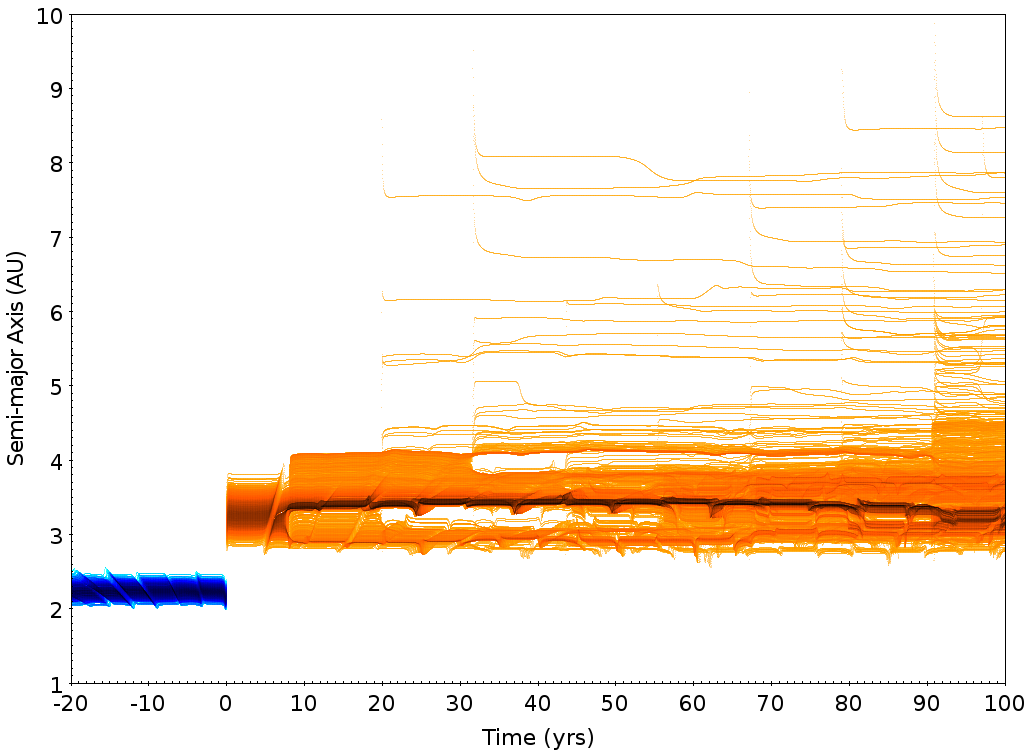}
         \caption{Semi-major axis variation}
         \label{fig:sub1}
     \end{subfigure}
     \\
     \begin{subfigure}[b]{\textwidth}
         \centering
         \includegraphics[width=0.45\textwidth]{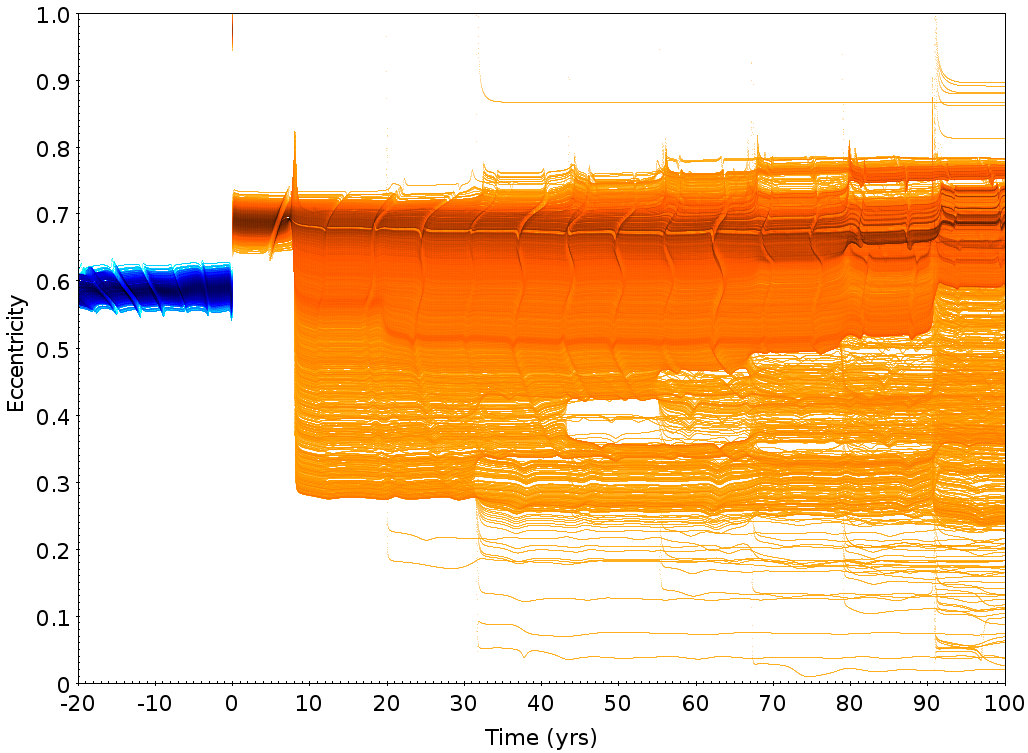}
         \caption{Eccentricity variation}
         \label{fig:sub2}
     \end{subfigure}
     \\
     \begin{subfigure}[b]{\textwidth}
         \centering
         \includegraphics[width=0.45\textwidth]{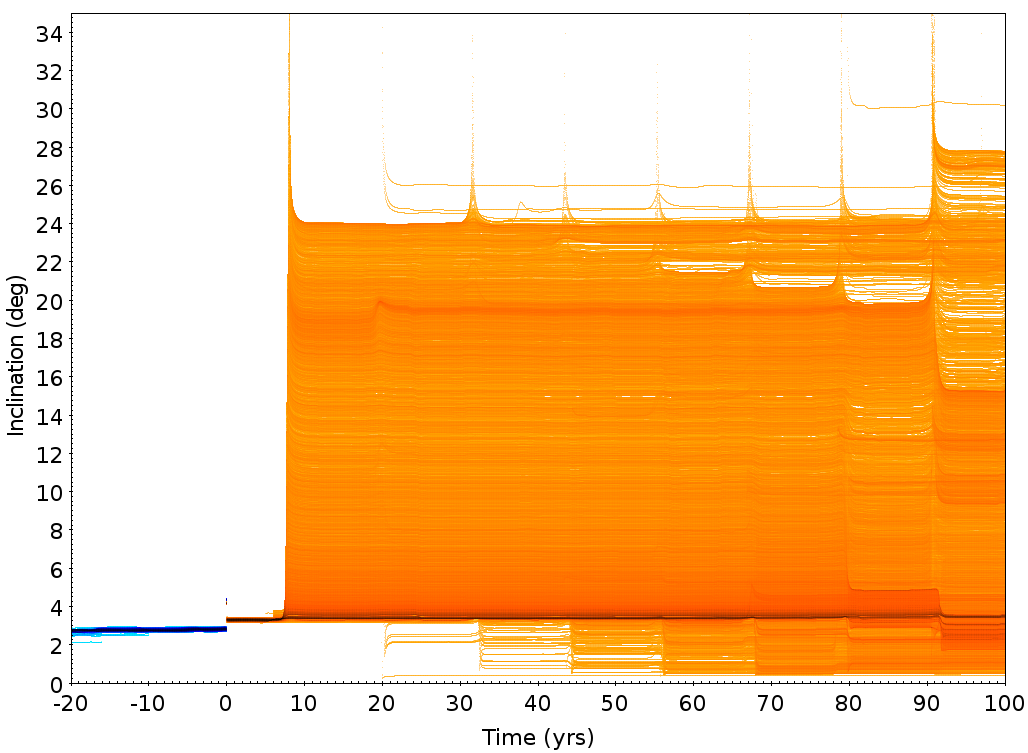}
         \caption{Inclination variation}
         \label{fig:sub2}
     \end{subfigure}
	\caption{Due to the grazing encounter with the Earth, the meteoroid from event DN170707\_01 was sent into a Jupiter intersecting orbit. On this new trajectory, the object will likely experience many close encounters with Jupiter over its lifetime. In these density plots, the blue and orange particles represent the meteoroid before and after the grazing encounter, respectively. The darker coloration is indicative of a higher particle density. The many possible close encounters with Jupiter manifest as discrete ``jumps" in the semi-major axis, eccentricity, or inclination. Over time the orbits tend to spread out due to numerous close encounters with Jupiter. Thus, the orbit of the meteoroid becomes less clear over a relatively short period of time.}
	\label{fig:aei_time}
\end{figure}

As seen in Fig.~\ref{fig:aei_time}, the well-constrained orbit prior to the close encounter with Jupiter rapidly spreads out in the orbital space. Following the likely meteoroid-Jupiter close encounter of 2025, the orbit of the meteoroid can only be treated statistically. The density plots in Fig.~\ref{fig:aei_time} show the evolution of the semi-major axis, eccentricity, and inclination of 10,000 test particles forward in time only 100\,years. Most of the particles stay together, indicated by the darker portions of the plot. However, as seen by the multiple jumps in values over time, the meteoroid is likely to have a plethora of close encounters with Jupiter over its lifetime in a JFC orbit, and every one of these encounters obscures the future of the object.  

\begin{figure}
	\centering
	\includegraphics[width=\textwidth,keepaspectratio]{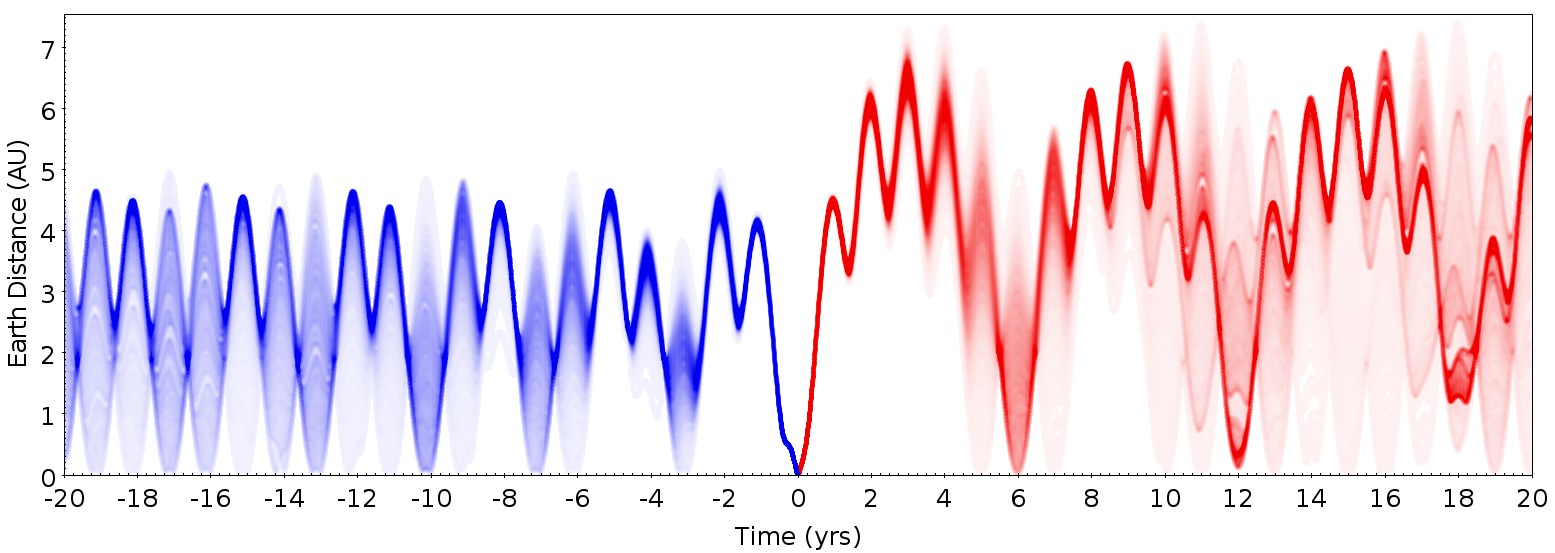}
	\caption{Plot of distance relative to the Earth over time. These 2 simulations were conducted with 5,000 particles and lasted 20\,years (forward and backward). The time on the x-axis is relative to the event epoch. The probability of an encounter with the Earth for the 20\,years before and after the fireball observation is extremely low. There is a $2.4\%$ probability of coming within 3 Hill radii and a $0.7\%$ probability having a 1 Hill radii encounter with the Earth within 20\,years before the grazing encounter. Meanwhile, there is a $1.4\%$ and a $0.5\%$ probability of the meteoroid encountering the Earth again within one and three Hill radii in the next 20\,years respectively. There is $1.1\%$ that the object will approach within 3 Hill radii in July 2023.}
	\label{fig:earth_d20}
\end{figure}

\paragraph{Close Encounters with Earth}
In order to determine the likelihood of future or previous close encounters with the Earth, two simulations with 5,000 particles were integrated both backward and forward 20\,years relative to the event (Fig.~\ref{fig:earth_d20}). During these simulations, outputs were collected at a higher frequency in order to accurately characterize all possible close encounters. The probability that there was an encounter with the Earth within three and one Hill radii within 20\,years prior to the grazing event was $2.4\%$ and $0.7\%$, respectively. Additionally, the probability that a future close encounter with the Earth will occur within the proceeding 20\,years after the grazing event is $1.4\%$ and $0.5\%$, respectively. Therefore, the probability of having the opportunity to telescopically observe this object as it re-approaches the Earth is very slim. The most likely time for this to occur is in mid-July 2023, but there is still only a $1.1\%$ chance that it will get within 3 Hill radii of the planet.

\subsection{Long-Term Simulations}
\begin{figure}
	\centering
    \includegraphics[width=\textwidth,keepaspectratio]{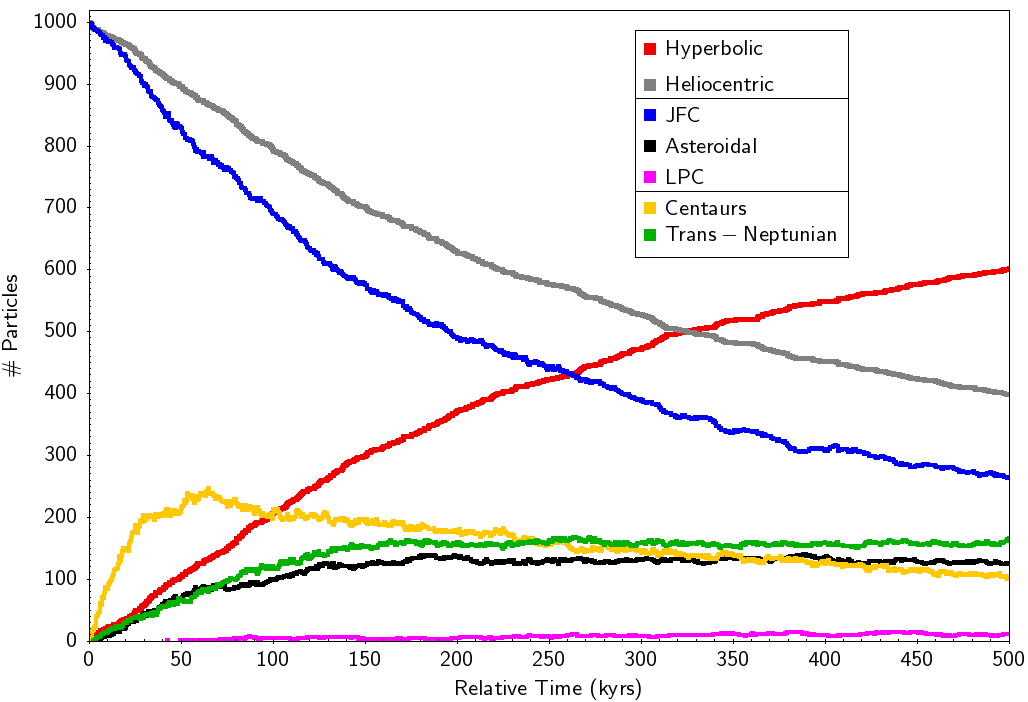}
	\caption{Plot showing the change in the orbital classification of the 1,000 particles in the forward integration of event DN170707\_01 for 500\,kyrs. The lines separating the labels in the legend group classifications together that are mutually exclusive (e.g. particles cannot be simultaneously hyperbolic and heliocentric). Over time, the likelihood that the meteoroid will have a close enough encounter to eject it from the Solar System increases. By the end of the simulation, $60.1\%$ of the particles are ejected, $27.3\%$ are still on JFC orbits, and $12.6\%$ have remained in the Solar System but have either gone onto long-period cometary or asteroid-like orbits. Many of the particles ($\sim20\%$) evolve onto Centaur and then trans-Neptunian orbits due to close-encounters with Jupiter.}
	\label{fig:lifetime_plot}
\end{figure}

\begin{figure}
	\centering
	\includegraphics[width=\textwidth,keepaspectratio]{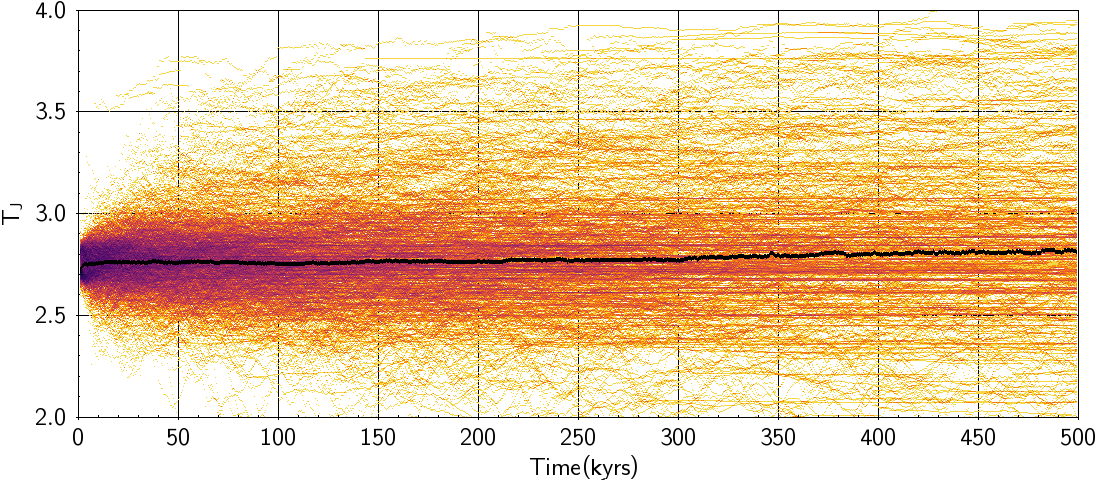}
	\caption{Tisserand's parameter variation of 1,000 particles integrated for $500\,kyrs$ post-grazing event for DN170707\_01. Particles that are ejected from the Solar System are removed. The meteoroid is likely to stay in a JFC orbit for an amount of time normal for a `natural' JFC object. The colouration in the plot is indicative of density - darker in areas of higher particle density and lighter in areas of lower particle density.}
	\label{fig:Tj_500ka}
\end{figure}

Further analysis using substantially longer integrations of test particles was performed in order to statistically characterize the meteoroid's future. The longest of these simulations was a forward integration of 1,000 test particles for 500\,kyrs. Over the course of the 500\,kyr forward integration, most of the particles ($60.1\%$) are eventually ejected from the Solar System, as expected (Fig.~\ref{fig:lifetime_plot}). The vast majority of the particles that remain in the Solar System (heliocentric orbits) stay in JFC orbits (as defined by the Tisserand's parameter) for the entire integration (Fig.~\ref{fig:Tj_500ka}). 

As seen in Fig.~\ref{fig:lifetime_plot}, there is an exponential decay in the number of particles in heliocentric and JFC orbits. The average dynamical lifetime for the particles in JFC orbits is approximately 200\,kyrs, which is shorter than the $\sim325$\,kyrs dynamical lifetime estimate for JFCs \citep{duncan2004dynamical,levisonduncan1997}. This is likely due to the initial post-grazing orbit, which has an aphelion very near the orbit of Jupiter. However, bodies in JFC orbits that display cometary features are more likely to have multiple $\leq0.1$\,AU encounters with Jupiter, reducing the orbital stability compared to asteroidal interlopers within the population \citep{fernandez2015jupiter}. Therefore, the JFC-orbit dynamical lifetime for the meteoroid is indistinguishable from a JFC from a more ``traditional" source region. In Fig.~\ref{fig:lifetime_plot}, the JFC, asteroidal, and LPC categories are solely determined by the particles' Tisserand's parameter. Whereas, the Centaur and trans-Neptunian objects are defined as having orbits between Jupiter and Neptune, and beyond the orbit of Neptune, respectively. This classification does lend itself to including some Centaurs and trans-Neptunian objects when counting the number of JFCs.

\begin{figure}
	\centering
    \includegraphics[width=\textwidth,keepaspectratio]{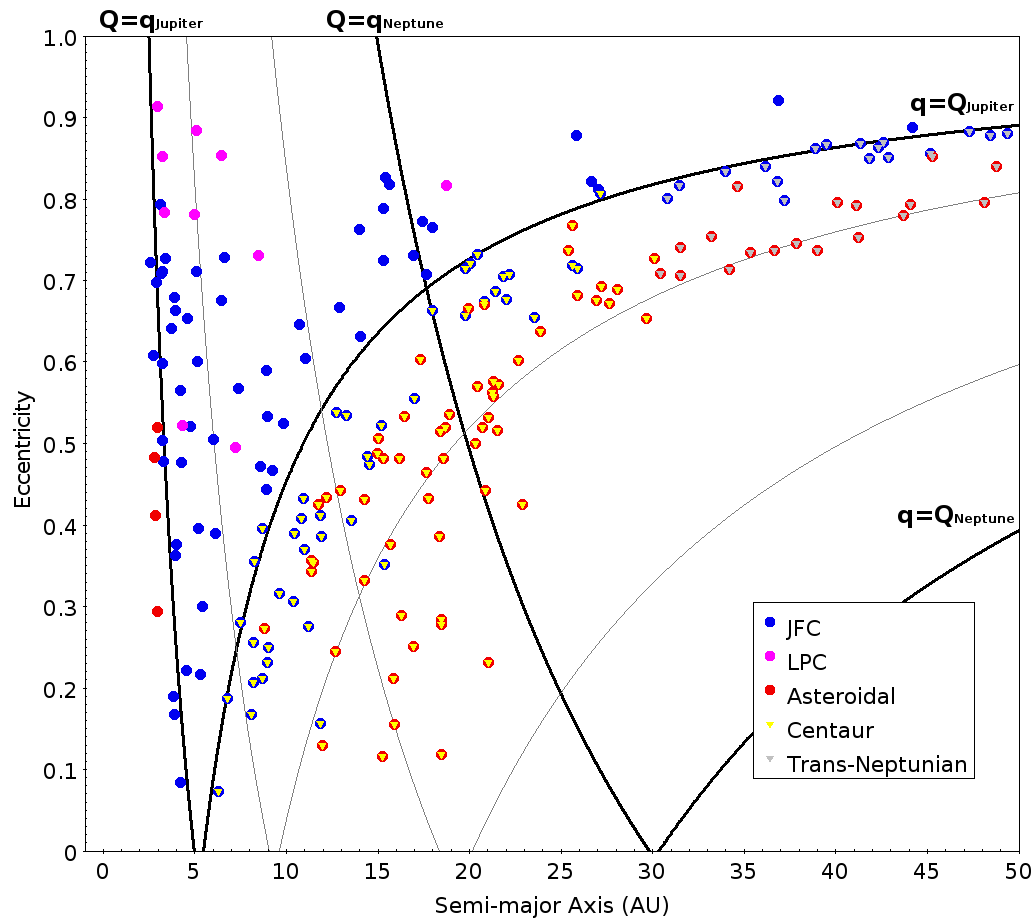}
	\caption{Semi-major axis vs eccentricity for the particles that are still gravatationally bound to the Sun after the forward integration of 1,000 particles for 500\,kyrs. The lines of equal perihelion and aphelion are plotted for Jupiter and Neptune in black, and in gray for Saturn and Uranus. Most of the remaining heliocentric particles ($68.4\%$) are in JFC-like orbits (blue) according to their Tisserand's parameter. However, a considerable number of particles ($29.1\%$) are considered ``asteroidal", according to their Tisserand's parameter with respect to Jupiter (red). Only a very small number of these are in the inner Solar System. A vast majority appear to be former JFC particles that have completely decoupled from Jupiter and have drifted onto Centaur and trans-Neptunian orbits due to planetary perturbations over time. The Centaur and trans-Neptunian particles can have either JFC or asteroidal like Tisserand's parameters ($T_{J}>2$), as shown by the colourisation in this plot. Also, about $2.5\%$ of the particles are categorized as ``long-period comets" despite their low semi-major axis because they are orbiting in retrograde orbits.}
	\label{fig:last_500ka}
\end{figure}

  A smaller fraction ($31.6\%$) of the test particles evolve onto asteroidal or long-period cometary orbits as defined by their Tisserand's parameter values (Fig.~\ref{fig:last_500ka}). The majority of asteroidal particles are determined to be in the outer Solar System. These are particles that originated from JFC space that were decoupled from Jupiter over time due to planetary perturbations and are now on Centaur or trans-Neptunian orbits. This portrays a similar mechanism alluded to in the concluding statements in  \citet{meech2004comet}, where they estimate there should be $\sim20$ objects of kilometer-size from the main-belt being scattered by Jupiter every million years in today's Solar System. The object discussed in this study differs in that it was gravitationally scattered by the Earth and then by Jupiter, resulting in the possible transfer of volatile-depleted inner Solar System material to the outer Solar System. 

\subsection{Analysis of Other Grazing Meteors/Fireballs} 

Within the current scientific literature, there have been in total ten grazing fireballs observed. However, in only six of these cases did the meteoroid survive the atmospheric passage and return to interplanetary space (Table~\ref{tab:othergrazers}). These grazing events demonstrate the orbital changes experienced by meteoroids that come very close to Earth. In most of these occurrences, the objects experience a significant change to their orbits. Although, this does not necessarily change them enough to be orbitally reclassified. For the first photographically observed grazing fireball, in October of 1990, a $10^{5} - 10^{6}$\,kg meteoroid in a higher inclination Apollo-type orbit with a JFC-like Tisserand's parameter was inserted into a lower-energy orbit with a $T_{J} > 3$. Thus, not only has a meteoroid with a more asteroid-like $T_{J}$ become more cometary due to close encounters, but the reverse has also been observed. It has been shown that using the Tisserand's parameter is a better metric to classify small Solar System bodies compared to the traditional arbitrary classification based on the orbital period \citep{carusi1987dynamical, levisonduncan1997}. Nevertheless, as shown in this study, small meter-sized objects occasionally experience close encounters with the Earth and have a sufficient orbital energy change to be reclassified even under this scheme. 


\begin{table}
\centering
\caption{Summary of six of the ten previous Earth-grazing meteors within scientific literature in which the meteoroid survived the passage through the atmosphere. Information omitted in the table was not included in the corresponding study.}
\begin{adjustbox}{width=0.95\paperwidth, center}
\begin{tabular}{lllllllll}
\toprule
Event Date     & Event Location            & Detection Method               & Initial Mass         & Orbit Before            & Orbit After             & $T_{J}$ Before & $T_{J}$ After & Reference          \\ \midrule
Aug. 10, 1972  & Western US and Canada     & satellite infrared radiometer  & $10^{5} - 10^{6}$\,kg & Amor                    & Apollo                  & 4.14           & 4.52          & \citet{ceplecha1979earth} \\
Oct. 13, 1990  & Czechoslovakia and Poland & photographic                   & $\sim$44\,kg           & Apollo                  & Apollo                  & 2.27           & 3.07          & \citet{borovicka1990comparison} \\
Sept. 23, 2003 & Ukraine                   & video                          &                      & Apollo                  & Apollo                  & 0.66           & 0.79          & \citet{kozaketal}  \\
Mar. 29, 2006 & Japan                     & video, photographic, telescope & $\sim$100\,kg         & JFC                     &                         & 2.85           &               & \citet{abe2006earth} \\
June 10, 2012  & Spain                     & video                          & 1.5 - 115\,kg         & Daytime $\zeta$-Perseid & Daytime $\zeta$-Perseid       & 3.31           & 4.04          & \citet{madiedo2016earth} \\
Dec. 24, 2014  & Spain                     & video                          &                      & Apollo                  &                         & 5.3            &               & \citet{moreno2016preliminary} \\
\bottomrule
\end{tabular}
\end{adjustbox}
\label{tab:othergrazers}
\end{table}

\paragraph{Other Grazing DFN Events}
If we consider an fireball event to be grazing simply when the initial slope of the trajectory is $<5\degree$ and travelled $>100$\,km through the atmosphere, in the four years since the DFN has being fully operational, we have observed $\sim1.2\%$ of the DFN dataset to be grazing events. Indicating that although somewhat uncommon, grazing events are not extremely rare. However, in most of the events detected, the meteoroid either does not survive the atmospheric passage or loses enough velocity to be incapable of re-entering interplanetary space.

\subsection{Implications and Further Research} 
Grazing fireballs indicate that meter-scale NEOs are occasionally inserted into categorically new orbits due to close encounters with the Earth, or indeed other planets. How effective this mechanism is for mixing material in the inner Solar System for small objects is still to be determined. Current work is being done to produce an artificial dataset of close encounters undetected by telescopes based upon the entire orbital dataset of the DFN (Shober et al., in prep.). This analysis will be extremely valuable to conclusively determine how significant this process is for small objects in the inner Solar System. If it is non-negligible, what populations in the near-Earth space may be more or less contaminated by genetically unrelated material, how significant are the orbit alterations, and what may this imply about where meteorites come from?

\section{Conclusions}
On July 7th, 2017, the Desert Fireball Network observed a $>1300$\,km long grazing fireball by ten of its high-resolution digital fireball observatories. The meteoroid transited the atmosphere for over 90\,seconds and reached a minimum height of 58.5\,km before returning to interplanetary space. This fireball is only matched by the notorious `Great Daylight Fireball of 1972', which penetrated to a very similar depth in the atmosphere but lasted $\sim9$\,seconds longer. As a result of the grazing encounter with the Earth, the meteoroid observed by the DFN underwent a natural slingshot maneuver in which it was transferred from an asteroidal Apollo-type orbit to a JFC orbit. Additionally, numerical integration of the object forward 500\,kyrs indicated that it will most likely stay in a JFC orbit for $\sim200$\,kyrs -- indistinguishable from any other JFC. Considering there are likely many small objects that go telescopically undetected that have close encounters with the Earth, there may be a non-negligible amount of meter-sized objects in modified orbits within the inner Solar System.


\section{Acknowledgements}
This work was funded by the Australian Research Council as part of the Australian Discovery Project scheme (DP170102529). SSTC authors acknowledge  institutional support from Curtin University.

This research made use of Astropy, a community-developed core Python package for Astronomy \citep{robitaille2013astropy}. Simulations in this paper made use of the REBOUND code which can be downloaded freely at http://github.com/hannorein/REBOUND \citep{2012A&A...537A.128R}. 

The authors would also like to thank David Clark for his assistance in creating animations of event DN170707\_01 and searching telescope surveys for images of the meteoroid before and after DFN observations. Additionally, we would like to thank Prof. J\"{u}rgen Oberst for being being extremely helpful and providing additional information about the grazing fireball observed over central-Europe in 2014.

\bibliography{skippy}

\bibliographystyle{abbrvnat}

\end{document}